\documentclass[pre,twocolumn,aps,eqsecnum]{revtex4}
\usepackage{epsfig}

\begin{document}
\begin{titlepage}

\title{Brittle-Ductile Transitions in a Metallic Glass}
\author{J.S. Langer}
\affiliation{Kavli Institute for Theoretical Physics, Kohn Hall, University of California,
        Santa Barbara, CA 93106}
\date{\today}
\begin{abstract}
Recent computational and laboratory experiments have shown that the brittle-ductile transitions in metallic glasses such as Vitreloy1 are strongly sensitive to the initial effective disorder (or ``fictive'') temperature.  Glasses with lower effective temperatures are weak and brittle; those with higher effective temperatures are strong and ductile. The analysis of this phenomenon presented here examines the onset of fracture at the tip of a slightly rounded notch as predicted by the shear-transformation-zone (STZ) theory of spatially varying plastic deformation.  The central ingredient of this analysis is an approximation for the dynamics of the plastic zone formed by  stress concentration at the notch tip. This zone first shields the tip but then breaks down suddenly producing a discontinuous transition between brittle and ductile failure, in semiquantitative agreement with the numerical and experimental observations.    
\end{abstract}
\maketitle

\end{titlepage}

\section{Introduction}
\label{Intro}

Two recent developments in fracture mechanics have interesting implications for materials theory. Specifically, the numerical simulations of amorphous crack-tip dynamics by Rycroft and Bouchbinder \cite{RB-12,VRB-16} and the related experimental results for metallic glasses by Ketkaew et al.\cite{SCHetal-18} both demonstrate that amorphous materials are embrittled by forming them with low densities of flow defects.   References \cite{RB-12,VRB-16} show that notch-like indentations are weak and brittle at low effective disorder temperatures and correspondingly low initial densities of shear-transformation zones (STZ's)\cite{FL,FL-11}; and that  they become stronger and more ductile at higher effective temperatures.  According to Ref.\,\cite{SCHetal-18} (see also \cite{SCHetal-13}), crack formation in metallic glasses is enhanced by decreasing their fictive temperatures.  That is, glasses are embrittled by quenching them slowly enough through their glass temperatures that they settle into states of relatively low disorder. Conversely, they remain tougher when quenched more quickly.  

Fracture toughness is a central issue in materials science that has long been addressed primarily by phenomenology.  However, we now have the STZ theory for amorphous plasticity \cite{FL,FL-11} and the thermodynamic dislocation theory for crystalline materials \cite{LBL-10,JSL-17a,JSL-19}, both of which are based on fundamental nonequilibrium statistical physics and have been tested by experiment in important but as-yet limited ways.  With \cite{RB-12} and \cite{SCHetal-18}, we have simulational and experimental results directly relevant to the brittle-ductile problem.  Thus the time seems ripe to look again at the basic theory of these phenomena and try to understand what is happening.  

Here I describe an attempt to interpret the results of \cite{RB-12} and \cite{SCHetal-18} analytically, and thus to obtain some basic understanding of these phenomena.  My strategy is to use an elliptical approximation to describe the tip of a notch in a sheet of material subject to an increasing, mode-I, opening stress.  My main assumption is that a crack is launched near this tip when the tensile stress, i.e. the negative pressure in its neighborhood, reaches some material-dependent threshold.    Rycroft and Bouchbinder \cite{RB-12} assume that cracks in metallic glasses are initiated by stress-induced cavitation events; but there are many other mechanisms that could be operative in other kinds of materials.  The critical stress for crack initiation will be one of the important system-dependent parameters in this theory. 

I start by considering simple Bingham plasticity (a special case of the STZ theory) with a linear increase in the rate of plastic deformation as a function of stress above a fixed yield stress; and I look at the onset of fracture near the tip of a notch where the rising stress is highly concentrated.  I find that both the elastic and plastic dynamics drive the tip to move forward and to sharpen.  Here I depart from the conventional wisdom that assumes plasticity always to be a blunting mechanism; but the sharpening effect is obvious from simple physics.  The growing concentrated transverse stress in front of the tip moves it forward, and sharpening occurs because the stress concentration is larger at the tip than behind it. This behavior will be shown mathematically in what follows. 

The Bingham analysis to be presented here tells us most  -- but not all -- of what we need to know about glassy fracture toughness. As will be seen, the Bingham solid undergoes a smooth transition from brittle to ductile behaviors as the relative strength of plastic versus elastic deformation is increased. When the plastic response is much slower than the loading rate, the fracture threshold is reached by the elastic forces alone and thus the fracture toughness is a relatively small constant as a function of loading rate.  This looks like brittle behavior.  

With increased plasticity or slower driving, a plastic boundary layer forms at the notch tip and partially shields it from the external stress, thus suppressing the elasticity-induced fracture.  Then, when the far-field stress exceeds the yield stress, the boundary layer expands rapidly and the tip stress grows suddenly, thus initiating ductile failure. This rapid expansion of the plastic zone is a well known feature of simple elasto-plastic theories (e.g. see \cite{BLLP-07}.) It plays a major role in the present theory.  But it does not cause a sharp transition between brittle and ductile behaviors in the Bingham model, at least not in the analysis described here.

The inclusion of STZ dynamics markedly changes this picture.  If the system has been quenched to a low effective temperature, then the work done by plastic deformation at the notch tip generates new STZ flow defects, increasing the local plastic deformation rate, and further increasing the STZ production rate.  This nonlinear instability eventually produces a sharp transition between brittle and ductile behaviors.  It is the central theme of this paper.  

Some mathematical elements of this fracture-toughness theory are described in Sec. \ref{Ellipse}.  Sections \ref{Bingham} and \ref{Bingham2} present the Bingham analysis; Sections \ref{Teff}  and \ref{Numerics}  describe the effective-temperature analysis and its predictions.  Section \ref{Questions} contains concluding remarks. Some mathematical details are provided in an Appendix.  
  
\section{Mathematical Preliminaries: The Elliptical Model}
\label{Ellipse}

Consider a plate of elasto-plastic material lying in the  $x,y$ plane and containing an elliptical hole.  The ellipse is highly  elongated in the $x$ direction and thus has sharp tips at its ends on the $x$ axis.  A mode-I stress $\sigma_{\infty}$ is imposed in the $y$ direction very far from the hole. If we assume symmetry about the $y$ axis, then this model is equivalent to a sharp notch with an opening stress  $\sigma_{\infty}$. 

My scheme is to use the elasto-plastic equations of motion to determine the behavior of this elliptical notch under steadily increasing values of $\sigma_{\infty}$.  There is an obvious difficulty.  We know that this shape does not remain elliptical under strong forcing; its irreversible motions must involve shape changes that cannot be described simply by time dependent values of the position and curvature of the tip.  To  minimize this difficulty, I focus only on the immediate neighborhood of the tip and look only at the early onset of plastic deformation there. By the end of this paper, we shall see important limitations of this strategy.   

The first step is to transform from Cartesian coordinates  $(x,y)$ to elliptical coordinates ($\rho$, $\theta$):
\begin{equation}
\label{rhoW}
x=W\left(\rho+{m\over\rho}\right)\,\cos\,\theta, ~~~y=W\left(\rho+{m\over\rho}\right)\,\sin\,\theta.
\end{equation}
Curves of constant $\rho$ are ellipses, and curves of constant $\theta$ are orthogonal hyperbolas. If we take the boundary of the elliptical hole to be at $\rho=1$, then the semi-major and semi-minor axes of the ellipse have lengths $W(1+m)$ and $W(1-m)$ respectively. Let $0<m<1$ so that the long axis of the ellipse lies in the $x$ direction, perpendicular to the applied stress, in analogy to a mode-I crack. 

To produce the long, thin ellipse, let $W$ become much larger than any other length in the system, and set $m\le 1$ so that the curvature at the tip, i.e. at $x=W(1+m)$, $y=0$, is large but finite.  Denote this curvature by ${\cal K}_{tip}$. Then a calculation to leading order in $1/\sqrt{W}$ yields
\begin{equation}
\label{m}
m \approx 1-2\,\epsilon;~~\epsilon\equiv\sqrt{1\over 2 \,{\cal K}_{tip} W}\,\ll 1,
\end{equation}
where $\epsilon$ will be the principal small parameter in this analysis.

The linearly elastic version of this problem has been solved by Muskhelishvili \cite{MUSK-63}.  His general results are summarized in the Appendix, Eqs. (\ref{sigmaeqn1} -\ref{devstrss}).  For our purposes, his most important formula is the expression for the deviatoric stress $s(\rho,\theta) \equiv s_{\theta\theta}(\rho,\theta) = - s_{\rho\rho}(\rho,\theta)$ given in Eq.(\ref{sigmaeqn2}), which can be used to derive an approximation for  $s(\rho,\theta)$  near the tip.  Let $\rho = 1 + \tilde x$, use the definition of $\epsilon$ in Eq.(\ref{m}), and assume that $\theta \ll \epsilon$.  I find:
\begin{equation}
\label{sapprox1}
s(\rho,\theta)\equiv  s(\tilde x,\theta)\cong {\sigma_{\infty} \epsilon^2\over (\epsilon + \tilde x)^3}\,\Bigl(1 - {2\,\theta^2\over \epsilon^2}\Bigr).
 \end{equation}
 
For $\tilde x = 0$ and $\theta = 0$, Eq.(\ref{sapprox1}) can be written
 \begin{equation}
 {s(0,0)\over s_y} = {\sigma_{\infty}\over s_y\,\epsilon} \equiv \psi \sqrt{\kappa},
 \end{equation}
 where 
 \begin{equation}
 \label{psidef}
 \psi \equiv {\sigma_{\infty}\over s_y}\sqrt{2W\over a};~~ \kappa \equiv a\, {\cal K}_{tip}. 
 \end{equation}
Here, $s_y$ is the plastic yield stress, which will play a prominent role in what follows, but which has been introduced here primarily for dimensional convenience.  Similarly, $a$ is a length scale of the order of magnitude of the initial tip radius, also included for dimensional reasons. (Unlike $\kappa$, $a$ is not a dynamical variable.) Thus, $\psi$ is a dimensionless measure of the stress intensity factor, where the applied stress $\sigma_{\infty}$ is amplified by the large factor $\sqrt{2 W/a}$.  

At this point, we must begin to pay attention to the plastic zone that forms ahead of the tip when the stress given by Eq.(\ref{sapprox1}) would exceed the plastic yield stress $s_y$.  This happens at a nonzero value of $\tilde x$, say $\tilde x_{max}$.  Within this zone,  where $\tilde x < \tilde x_{max}$, Eq.(\ref{sapprox1}) is not valid.  For $\theta = 0$,  
\begin{equation}
\label{xmax}
\tilde x_{max} = \epsilon\,[[\nu - 1]],~~~ \nu \equiv \Bigl({\sigma_{\infty}\over \epsilon s_y}\Bigr)^{1/3} = (\psi\sqrt{\kappa})^{1/3}.
\end{equation}  
The onset of plastic deformation at the tip occurs when $\nu = 1$; that is, when $\psi = \kappa^{-1/2}$.  

I have introduced a notation here that will be convenient in much of what follows.  For any quantity $f$, the double square brackets mean that $[[f]] = f$ if $f \ge 0$ and $[[f]] = 0$ otherwise.       

It is important to understand the significance of the quantity $[[\nu - 1]]$.  According to Eq.(\ref{rhoW}), the position of the tip on the  $x$ axis is 
\begin{equation}
x_0 = W\,(1 + m) \cong 2 W (1- \epsilon) ,
\end{equation}
and the front edge of the plastic zone is at
\begin{eqnarray}
\nonumber&& x_{max} = W\,\Bigl(1 + \tilde x_{max} + {1- 2\epsilon\over 1 + \tilde x_{max}}\Bigr)~~~ \cr\\&&\cong x_0 +{1\over {\cal K}_{tip}} [[\nu - 1]].
\end{eqnarray} 
Thus, $[[\nu - 1]]= {\cal K}_{tip}(x_{max}-x_0)$ is the thickness of the plastic zone in units of the radius of curvature at the tip. 

\section{Bingham Elastoplasticity}
\label{Bingham}

The next stage of this investigation is to develop an analytic  approximation for elasto-plastic dynamics near the notch tip  using only the Bingham model.  

For simplicity, assume that the material is incompressible.  Also assume hypo-elasto-plasticity (additive decomposition of elastic and plastic rates of deformation). These assumptions imply that the diagonal elements of the rate-of-deformation tensor have the form
\begin{eqnarray}
\label{RODT}
\nonumber 
 && D_{\theta\theta}(\rho,\theta) = - D_{\rho\rho}(\rho,\theta)\cr\\&&\equiv D(\rho,\theta)\cong {1\over 2 \mu}{ds(\rho,\theta)\over dt}+ D^{pl}(\rho,\theta),
 \end{eqnarray}
 where $s(\rho,\theta) = s_{\theta\theta}(\rho,\theta)= -s_{\rho\rho}(\rho,\theta)$ is the deviatoric stress.  Bingham plasticity, with a yield stress $s_y$ and a constant plastic rate factor $1/ \tau_{pl}$, means that 
 \begin{equation}
 \label{BINGHAM}
 D^{pl}(\rho,\theta) \cong {1\over \tau_{pl}}\,[[{s(\rho,\theta)\over s_y}-1]] ,
\end{equation}
where the double square brackets mean the same thing that they did when introduced in Eq.(\ref{xmax}). For these purposes, we do not need to consider changes in direction of the stress field or even its values at large distances away from the $x$-axis. Assume that the important behavior is controlled by the elasto-plastic dynamics immediately ahead of the notch tip. 

The next question is how to evaluate the stress $s(\rho,\theta)$ for values of $\rho$ and $\theta$ inside the plastic region.  About a decade ago, my colleagues and I \cite{BLLP-07}  considered STZ elasto-plasticity in the neighborhood of an expanding circular hole, where the problem could be solved analytically because variations in the size of the hole and in the neighboring elasto-plastic fields occur only in the radial direction.  Our stated motivation was to gain some insight regarding the fracture problem.  I shall use two ideas from \cite{BLLP-07}, the first being a boundary-layer approximation, and the second a circular approximation for the  stress at the tip of the notch. 

The boundary-layer approximation for Eq.(\ref{BINGHAM}), at and just ahead of the tip, is:
\begin{equation}
\label{Dpl1}
D^{pl}(\rho,\theta)\equiv D^{pl}(\tilde x,\theta) \cong {1\over\tau_{pl}}\, [[{s(\tilde x,0)\over s_y} - 1]]\,\Bigl(1- {2 \,\theta^2\over \epsilon^2}\Bigr)~~~~~
\end{equation}
where, for $0 < \tilde x <  \tilde x_{max}$  and $ s(0,0)> s_y $,
\begin{equation}
\label{linear-s}
{s(\tilde x,0)\over s_y}- 1 \cong ({s_0\over s_y} - 1)\Bigl(1- {\tilde x\over \tilde x_{max}}\Bigr).
\end{equation}
That is, $s(\tilde x,0)$ is approximated by a linear function of $\tilde x$ across the boundary layer; and $s_0 =  s(0,0)$ is a time dependent boundary stress yet to be determined.  This  kind of approximation worked well in \cite{BLLP-07}; I shall use it throughout this paper.  Note also that, for mathematical consistency in Eq.(\ref{Dpl1}), I have kept only the lowest order correction in the angle $\theta$, moving that dependence outside the double square brackets. The angular dependence of the boundary layer is a higher-order correction in the limit of small $\epsilon$ and $\theta$.  

We now can use Eqs.(\ref{Drhorho}) and (\ref{Dthetatheta}) in the Appendix to express the rate of deformation tensor ${\cal D}$ in terms of the material velocities  $v_{\rho}$ and $v_{\theta}$ near the crack tip, and thus use Eq.(\ref{RODT}) to write equations of motion for those velocities.  Using the same approximations for small $\tilde x$ and small $\theta$ used above, I find
\begin{eqnarray}
\label{Drhorho2}
\nonumber
 {\cal D}_{\rho\rho}&&\approx {1\over 2 \epsilon W}\,\Bigl[{\partial v_{\rho}\over \partial\tilde x}+ {\partial v_{\theta}(0)\over\partial \theta}\,{\theta^2\over \epsilon^2}\Bigr]\,\Bigl(1 - {\theta^2\over 2\,\epsilon^2}\Bigr)\cr\\&& = -\, D(\tilde x,\theta);
\end{eqnarray}
and
\begin{eqnarray}
\label{Dthetatheta2}
\nonumber
{\cal D}_{\theta\theta}
&&\approx {1\over 2 \epsilon W}\,\Bigl[{\partial v_{\theta}\over \partial\theta}+ {v_{\rho}\over\epsilon}\,\Bigl(1- {\theta^2\over \epsilon^2}\Bigr)\,\Bigr]\Bigl(1 - {\theta^2\over 2\,\epsilon^2}\Bigr)\cr\\&& = + \,D(\tilde x,\theta),
\end{eqnarray}
where 
\begin{equation}
D(\tilde x,\theta) \cong D_0(\tilde x) \Bigl(1- {2 \,\theta^2\over \epsilon^2}\Bigr),
\end{equation}
and
\begin{equation}
\label{D0}
D_0(\tilde x) = {1\over 2\mu} \dot s(\tilde x,0)  + D^{pl}(\tilde x,0).
\end{equation}
In evaluating $D_0(\tilde x)$,  use Eq.(\ref{linear-s}) for the stress inside the plastic region, and assume that the time derivative of $s(\tilde x)$ is adequately approximated simply by taking the time derivative of $s_0$ in that equation.  Outside the plastic region,  use Eq.(\ref{sapprox1}) and take the time derivative of $\sigma_{\infty}$.  

\begin{figure}[h]
\begin{center}
\includegraphics[width=\linewidth] {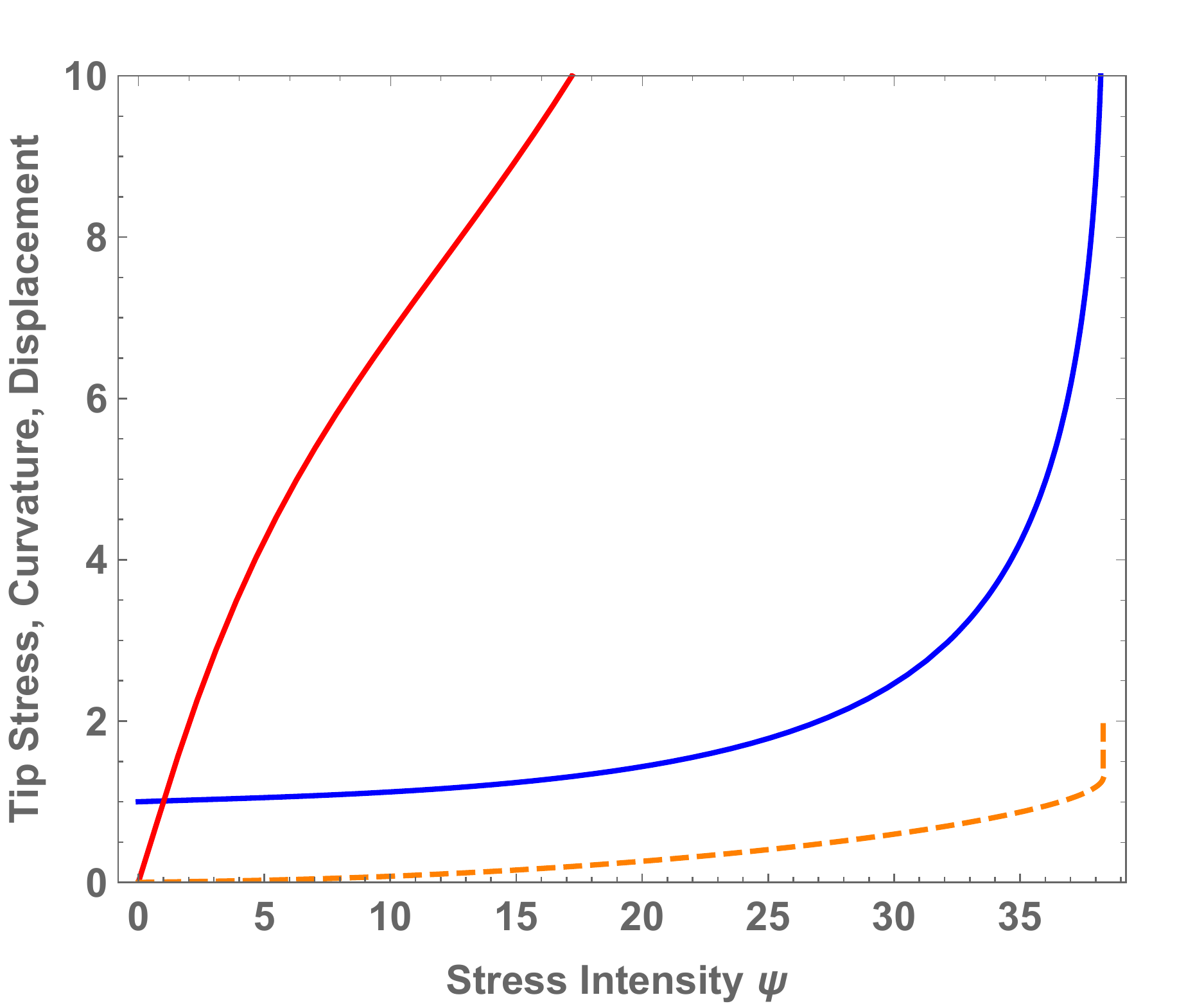}
\caption{Bingham model with $\xi = 0.003$.  The curves are, from top to bottom, the dimensionless tip stress $\tilde s_0$, the curvature $\kappa$, and the displacement $\tilde u_{tip}$ as functions of the dimensionless stress intensity factor $\psi$. }   \label{BD3-Fig-1}
 \end{center}
\end{figure}

Set $\theta = 0$ in Eq.(\ref{Drhorho2}), and use Eq.(\ref{D0}) to compute the tip velocity:
\begin{eqnarray}
\label{vtip}
\nonumber
&& v_{tip} = v_{\rho}(0)=-\int_0^{\infty} d\tilde x\,{dv_{\rho}\over d\tilde x}= 2\epsilon\,W \int_0^{\infty} d\tilde x\, D_0(\tilde x)\cr\\&&= \nonumber{a\over 2\kappa}(\bar\nu -1)\Bigl({1\over 2\mu} \dot s_0 + {1\over \tau_{pl}} [[{s_0\over s_y}-1]]\Bigr)\cr\\&& ~~~~~+ \Bigl({a\over 2\, \kappa\,\bar\nu^2}\Bigr) \Bigl({s_y\over 2\mu}\Bigr)\Bigl({\dot\sigma_{\infty}\over \epsilon s_y}\Bigr).
\end{eqnarray}
The final result shown here is obtained by integrating separately over the plastic zone ($0 <\tilde x <\tilde x_{max}$) and the  elastic region ($\tilde x_{max}<\tilde x <\infty$) . The quantity $\nu$ is defined in Eq.(\ref{xmax}) as a function of the stress-intensity factor and the tip curvature.  The related quantity $\bar\nu$ is defined so as to distinguish contributions from inside and outside the plastic region in the integrals over $\tilde x$: $\bar\nu \equiv \nu$ if $\nu\ge 1$ and $\bar\nu\equiv 1$ if $\nu< 1$.

Because the tip curvature $\kappa$ has become a time-dependent dynamical variable in these equations, we need an equation of motion for it.  Start with the geometric formula \cite{JSL-87}
\begin{equation}
\label{Ktip}
-{\dot{\cal K}_{tip}\over {\cal K}^2_{tip}} =  v_{tip} + {1\over 2\,{\cal K}_{tip}\,W}\,{\partial^2 v_{\rho}\over \partial\,\theta^2}\Bigr|_{\theta = 0}.
\end{equation}
To evaluate this expression, it is useful to define 
\begin{equation}
v_{\rho}(\tilde x,\theta) \equiv v_0(\tilde x)+v_2(\tilde x)\, {\theta^2\over\epsilon^2},
\end{equation}
so that Eq.(\ref{Ktip})  becomes
\begin{equation}
- {\dot \kappa\over \kappa} = {\kappa\over a}\,\bigl(v_0(0) + 2\,v_2(0)\bigr).
\end{equation}

\begin{figure}[h]
\begin{center}
\includegraphics[width=\linewidth] {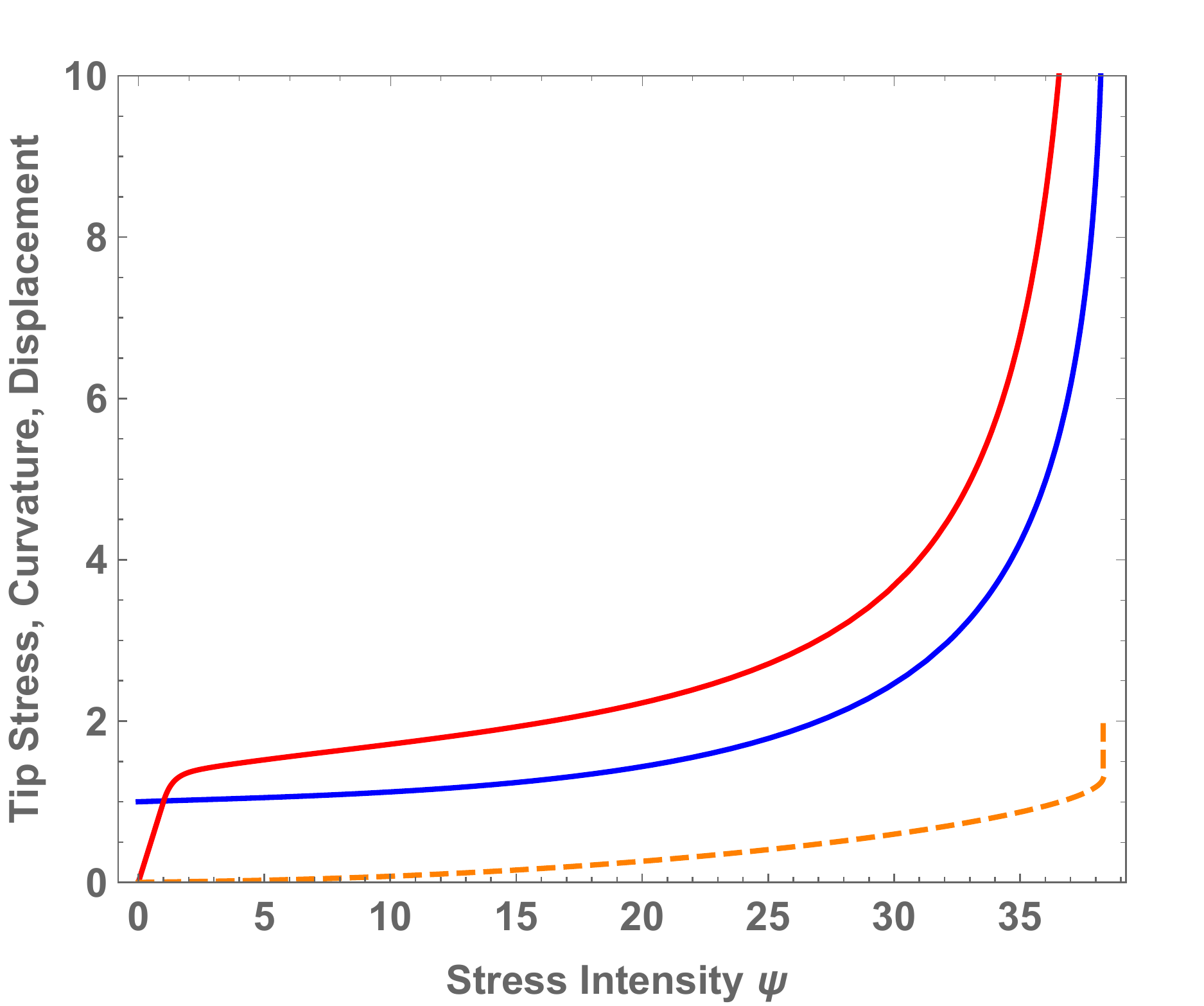}
\caption{Bingham model with $\xi = 0.03$.  The curves are, from top to bottom, the dimensionless tip stress $\tilde s_0$, the curvature $\kappa$, and the displacement $\tilde u_{tip}$ as functions of the dimensionless stress intensity factor $\psi$. }   \label{BD3-Fig-2}
 \end{center}
\end{figure}

Now use Eq.(\ref{Dthetatheta2}) at $\theta=0$ to write 
\begin{equation}
\Bigl({\partial v_{\theta}\over \partial\theta}\Bigr)_{\theta=0} =- {v_0(\tilde x)\over\epsilon} + 2\,\epsilon\,W D_0(\tilde x),
\end{equation}
and insert this into Eq.(\ref{Drhorho2}).  Collecting terms proportional to $\theta^2/\epsilon^2$, I find 
\begin{equation}
{d v_2\over d \tilde x} = \epsilon\,W\,D_0(\tilde x) +{v_0(\tilde x)\over\epsilon}.
\end{equation}
Then, using
\begin{equation}
\label{v0eqn}
{d v_0\over d \tilde x} = -2\epsilon\,W\,D_0(\tilde x)
\end{equation}
and combining terms, I find
\begin{eqnarray}
\label{kappadot}
\nonumber
{\dot \kappa\over \kappa}&=&{2\over \epsilon^2}\int_0^{\infty} \tilde x d\tilde x D_0(\tilde x)~~~~~~\cr\\&=& \nonumber {(\bar\nu -1)^2\over 3}\Bigl({1\over 2\mu} \dot s_0 + {1\over \tau_{pl}} [[{s_0\over s_y}-1]]\Bigr)\cr\\&&~~~~~~~~~+ 2\,\Bigl({2\bar\nu -1\over \bar\nu^2}\Bigr) \Bigl({s_y\over 2\mu}\Bigr)\Bigl({\dot\sigma_{\infty}\over \epsilon s_y}\Bigr).
\end{eqnarray}

With equations of motion for the tip position and curvature, it remains to find an equation of motion for the tip stress $s_0$. It is here that I shall use a circular approximation, similar to but not the same as the ones used in \cite{VRB-16} and \cite{BLLP-07}.  Consider a pair of concentric rings in a circular geometry with radial variable $r$ and a radial rate of deformation $v(r)$. The inner ring has a radius $R$ equal to the tip radius $a/\kappa$; and the outer ring is at the boundary of the plastic zone, thus at $R_1 = \bar\nu R$.  The analogs of the equations of motion, Eqs.(\ref{Drhorho2}) and (\ref{Dthetatheta2}), are 
\begin{equation}
{\partial v\over \partial r} + {v\over r} = 0,
\end{equation}
and
\begin{equation}
\label{dvdr2}
- {\partial v\over \partial r} + {v\over r} =2\,\Bigl({\dot s\over 2\,\mu} + {1\over \tau_{pl}}[[{s\over s_y} - 1]]\Bigr). 
\end{equation}
The first of these equations is the statement of incompressibility, which implies that $v(r) = R\,\dot R/r$.  If we make the boundary-layer approximation analogous to Eq.(\ref{linear-s}),
\begin{equation}
{s(r)\over s_y} \cong 1 + \Bigl({s_0\over s_y}- 1\Bigr)\,{R_1 - r\over R_1 - R},~~~R < r < R_1,
\end{equation}
 then we can integrate (\ref{dvdr2}) to find
\begin{equation}
\label{Rdot}
{\dot R\over R} - {\dot R_1\over R_1} = \Bigl({\dot s_0\over 2\,\mu} + {1\over \tau_{pl}}[[{s_0\over s_y} - 1]]\Bigr)\,\lambda(\bar\nu)
\end{equation}
where, using $R_1/R = \bar\nu$,
\begin{equation}
\lambda(\bar\nu) = 2\,\int_R^{R_1} {dr\over r}\Bigl({R_1 - r\over R_1-R}\Bigr)={2 \,\bar\nu\over \bar\nu-1}\,\ln \bar\nu - 2.
\end{equation}
Finally, use the expression for $v_{tip}$ in Eq.(\ref{vtip}) to evaluate $\dot R$, integrate Eq.(\ref{v0eqn}) to evaluate $\dot R_1$, insert these expressions into the left-hand side of Eq.(\ref{Rdot}), and solve for $\dot s_0$.  The resulting equation of motion for the tip stress is
\begin{equation}
\label{s0dot}
{\dot s_0\over 2\mu} = - {1\over \tau_{pl}}[[{s_0\over s_y} - 1]] + \Bigl({s_y\over 2\mu}\Bigr)\Bigl({\dot\sigma_{\infty}\over \epsilon s_y}\Bigr)\,\Lambda(\bar\nu),
\end{equation}
where 
\begin{equation}
\label{Lambdadef}
\Lambda(\bar\nu)= {\bar\nu -1\over 2\,\lambda(\bar\nu) - \bar\nu +1} = {(\bar\nu -1)^2\over 4\,\bar\nu \,\ln \bar\nu - (3 + \bar\nu)(\bar\nu -1)}.~~~~~~
\end{equation} 
Despite appearances, $\Lambda(\bar\nu)$ is continuous in both value and slope at the onset of plasticity at $\bar\nu =1$.  Importantly, it diverges at $\bar\nu \cong 5$ describing -- but only  approximately -- the sudden expansion of the plastic zone and rapid unshielding of the notch tip that occurs when the far-field stress exceeds the yield stress.    

\section{Solutions of the Bingham Equations}
\label{Bingham2}

Equations (\ref{vtip}), (\ref{kappadot}) and (\ref{s0dot}) provide a mathematically complete statement of the Bingham problem.  It will be useful to restate them in dimensionless form using variables introduced in Sec. \ref{Ellipse}.  

Let the dimensionless stress intensity factor $\psi = (\sigma_{\infty}/s_y)\sqrt{2W/a}$ be the principal independent variable, increasing linearly in time and thus serving as a time-like quantity.  Therefore $\dot\psi =(\dot\sigma_{\infty}/s_y)\sqrt{2W/a} \equiv 1/\tau_{ex}$ is a constant, and $1/\tau_{ex}$ is the external driving rate.  Then, $\dot\sigma_{\infty}/s_y\,\epsilon = \sqrt{\kappa}/\tau_{ex}$.  Measure stresses in units of $s_y$, so that $\tilde s_0\equiv s_0/s_y$; and define the dimensionless constant $c_0 \equiv s_y/2\,\mu$. ($c_0 \cong 0.01$ for Vitreloy1.) Define the ratio of time scales to be $\xi \equiv \tau_{ex}/\tau_{pl}$.  (Note that this $\xi$ is not exactly the same as the $\xi$ defined in \cite{VRB-16}.).  Importantly, define the critical failure stress in units of $s_y$ to be $\tilde s_c$.  According to \cite{RB-12,VRB-16}, $\tilde s_c \cong 4.5$ for Vitreloy1.

\begin{figure}[h]
\begin{center}
\includegraphics[width=\linewidth] {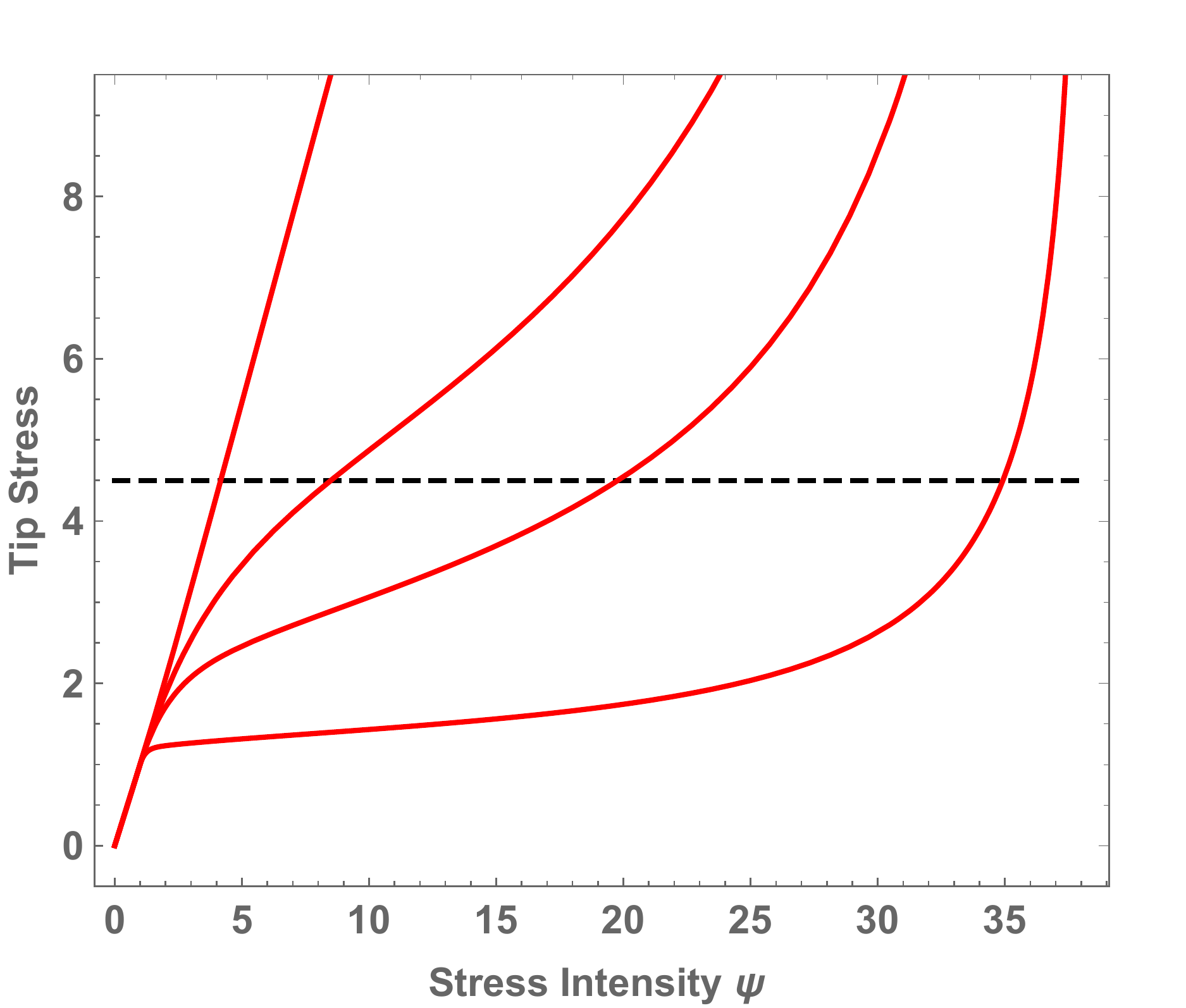}
\caption{Bingham model.  The curves are the dimensionless tip stresses $\tilde s_0$ as functions of the stress intensity factor $\psi$ for $\xi = 0.001,\,0.005,\,0.01,\,{\rm and}\,\,0.05$ from left to right. The horizontal line is at  $\tilde s_c = 4.5$.}   \label{BD3-Fig-3}
 \end{center}
\end{figure}
\begin{figure}[h]
\begin{center}
\includegraphics[width=\linewidth] {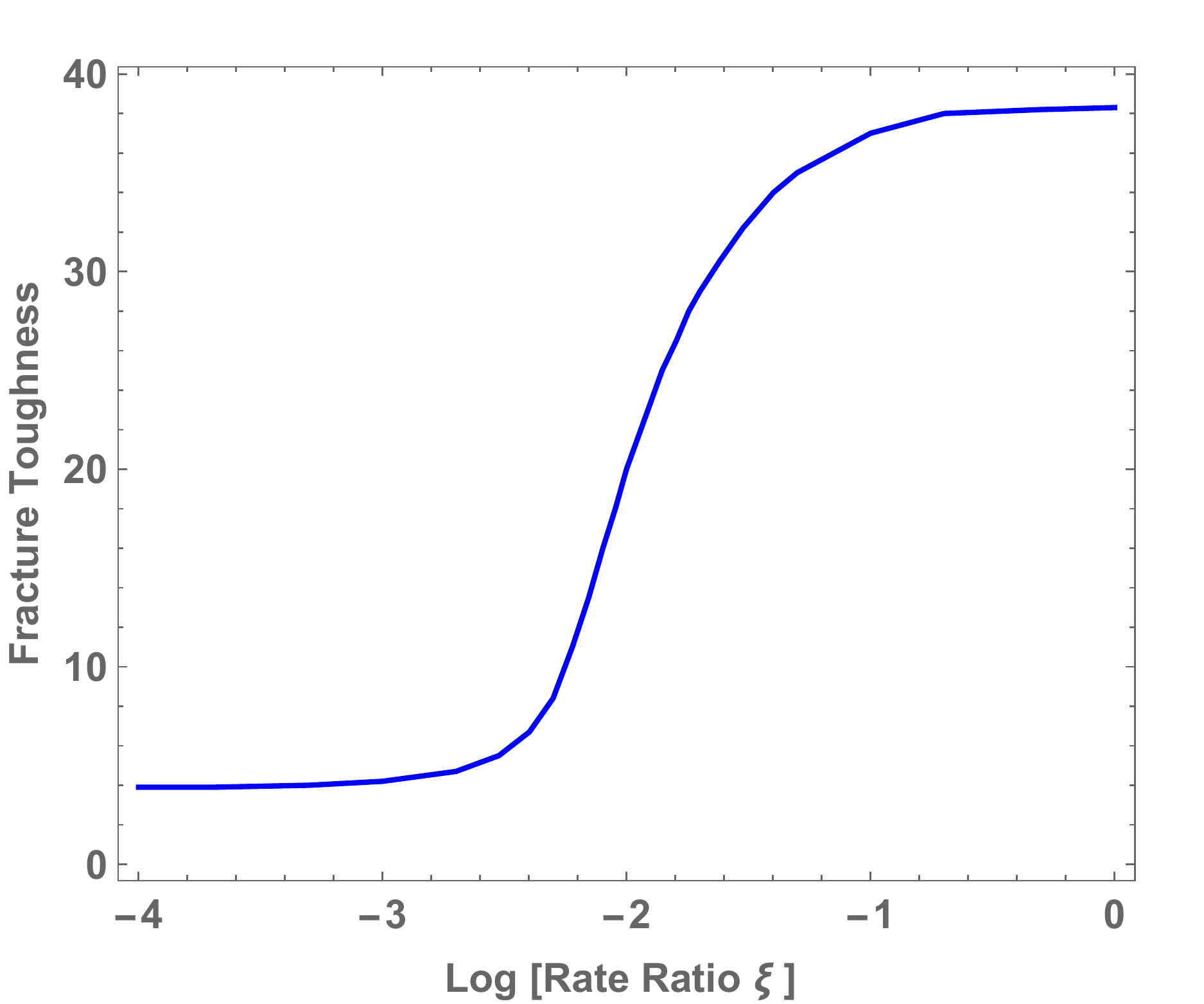}
\caption{ Dimensionless fracture toughness as a function of ${\rm Log}_{10} (\xi)$ for the Bingham model. }   \label{BD3-Fig-4}
 \end{center}
\end{figure}
Let $\tilde u_{tip}(\psi)$ be a dimensionless tip displacement for which $d \tilde u_{tip}/d\psi = (\tau_{ex}/a)\, v_{tip}$.  Then Eq.(\ref{vtip}) becomes an equation of motion for $\tilde u_{tip}$: 
\begin{equation}
\label{utip}
{d \tilde u_{tip}\over d\psi} = {\bar\nu -1\over 2\,\kappa} \Bigl(c_0 {d \tilde s_0\over d\psi} + \xi [[\tilde s_0 -1]]\Bigr) + {c_0\over 2 \bar\nu^2 \sqrt{\kappa}}.
\end{equation}
Similarly, Eq.(\ref{kappadot}) becomes 
\begin{equation}
{1\over \kappa}{d \kappa\over d\psi} = {(\bar\nu -1)^2\over 3} \Bigl(c_0 {d \tilde s_0\over d\psi} + \xi [[\tilde s_0 -1]]\Bigr) + 2 c_0 \sqrt{\kappa}\Bigl({2 \bar\nu -1\over  \bar\nu^2 }\Bigr);~~~
\end{equation}
and the tip-stress equation is:
\begin{equation}
\label{tipstress}
{d \tilde s_0\over d\psi} = - {\xi\over c_0} [[\tilde s_0 -1]] + \sqrt{\kappa}\, \Lambda(\bar\nu),
\end{equation}
where $\Lambda(\bar\nu)$ is defined in Eq.(\ref{Lambdadef}). 
Also, 
\begin{equation}
\bar\nu(\psi) = \cases{(\psi \sqrt{\kappa(\psi)})^{1/3},&if $ \psi \sqrt{\kappa(\psi)}>1,$\cr ~~~~~~~1,&otherwise.}
\end{equation}

Figures \ref{BD3-Fig-1} and \ref{BD3-Fig-2} show graphs of, from top to bottom, the dimensionless tip stress $\tilde s_0$ (red), the tip curvature $\kappa$ (blue), and the tip displacement $\tilde u_{tip}$ (orange dashed) as functions of the steadily increasing, dimensionless stress intensity factor $\psi$.  In the first figure $\xi = 0.003$; in the  second $\xi = 0.03$.  The first situation looks brittle; the tip stress rises almost linearly with the  applied stress and reaches its critical value for fracture, ($\tilde s_c\cong 4.5$) at $\psi \cong 5$. Both $\kappa$ and  $\tilde u_{tip}$ diverge at a much larger value of $\psi$ where the system theoretically would undergo rapid plastic failure; but a crack has been launched elastically before the system reaches that state.  

In Figure \ref{BD3-Fig-2}, where the plasticity strength $\xi$ is stronger by a factor of ten, the tip becomes shielded by a plastic boundary layer almost immediately as soon as the tip stress reaches the yield stress ($\tilde s_0 \cong 1$), and the graph of $\tilde s_0(\psi)$ bends over smoothly but abruptly.  As a result, the system undergoes ductile failure at a considerably larger value of $\psi$. 

Figure \ref{BD3-Fig-3} shows four $\tilde s_0(\psi)$ curves for  $\xi = 0.001, 0.005, 0.01, {\rm and} \,\,0.05$ along with a dashed line at $\tilde s_0 = \tilde s_c = 4.5$.  That line intersects the $\tilde s_0(\psi)$ curves at the corresponding fracture-toughness values of $\psi$. The full range of those fracture-toughness values as a function of $\xi$ is shown by the semi-log plot in Fig.\ref{BD3-Fig-4}.  This is the advertised smooth brittle-ductile transition for the Bingham model.

\section{Effective Temperature Dynamics}
\label{Teff}

To make contact with the Rycroft-Bouchbinder simulations \cite{RB-12}, we must introduce the space and time dependent effective temperature $\chi$ that determines the local density of flow defects, that is, the STZ's.  The basic assumption is that the plastic deformation rate is proportional to this density which, in turn, is proportional to an effective thermal activation factor. I write this modified rate factor in the form:
\begin{equation}
{1\over \tau_{pl}}\, e^{- e_Z/\chi(\theta,t)}\,e^{e_Z/\chi_{\infty}}
\end{equation}
where $e_Z$ is the STZ formation energy. The first factor, $1/\tau_{pl}$, is the same as the one introduced in Eq.(\ref{BINGHAM}) to describe Bingham plasticity.  The second is the STZ activation factor, and the last term adjusts that factor so that, in the steady-state limit $\chi \to \chi_{\infty}$, we recover the Bingham result.  

The effective temperature $\chi$ needs to be evaluated here only on the surface of the crack tip.  Thus, I modify Eq.(\ref{Dpl1}) to read
\begin{eqnarray}
\label{Dpl2}
&&\nonumber\tilde D^{pl}(\tilde x,\theta) = {1\over \tau_{pl}} \, e^{- e_Z/\chi(\theta,t)}\,e^{e_Z/\chi_{\infty}}\cr\\&&\times  [[{s(\tilde x,0)\over s_y} - 1]]\,\Bigl(1- {2 \,\theta^2\over \epsilon^2}\Bigr).
\end{eqnarray}
Let
\begin{equation}
\chi(\theta,t) = \chi(t) - \gamma(t)\,{\theta^2\over \epsilon^2};~~\gamma(t) = - {\epsilon^2\over 2}{\partial^2 \chi\over \partial\, \theta^2}\Big|_{\theta=0},
\end{equation}
and $\chi(t) \equiv \chi(\theta =0,t)$.  Then
\begin{equation}
e^{- e_Z/\chi(\theta,t)}\cong e^{- e_Z/\chi(t)}\Biggl(1 - {e_Z\,\gamma(t)\,\theta^2\over  \chi^2(t)\,\, \epsilon^2}\Biggr);
\end{equation}
\noindent and
\begin{eqnarray}
\label{Dpl3}
&&\nonumber\tilde D^{pl}(\tilde x,\theta) \cong  {1\over \tau_{pl}}\, e^{- e_Z/\chi(t)}\,e^{e_Z/\chi_{\infty}}\cr\\&& \times [[{s(\tilde x,0)\over s_y} - 1]] \Biggl[1- {\theta^2\over \epsilon^2}\Biggl(2 + {e_Z \gamma(t)\over \chi^2(t)}\Biggr)\Biggr].
\end{eqnarray}

We next need equations of motion for $\chi(t)$ and $\gamma(t)$. The basic equation of motion for $\chi(\theta,t)$ has the form
\begin{equation}
\label{chidot}
c_{e\!f\!f}\, \dot\chi(\theta,t) = \tilde D^{pl}(0,\theta)\, \tilde s(0,\theta) \Bigl[1 - {\chi(\theta,t)\over \chi_{\infty}}\Bigr].
\end{equation}
This is the effective heat-flow equation that has been conventional in STZ theory.  $c_{e\!f\!f}$ is the effective specific heat; the product $\tilde D^{pl}\, \tilde s$ is the rate at which power is delivered to the tip region by the plastic deformation; and $\chi_{\infty}$ is the steady-state value of $\chi$.  Eq.(\ref{Dpl3}) tells us what to use for $\tilde D^{pl}(0,\theta)$ here; and $\tilde s(0,\theta) \approx s_y (1- 2 \theta^2/\epsilon^2)$ is accurate enough for this purpose.  Inserting these ingredients into Eq.(\ref{chidot}) and setting $\theta = 0$, we find
\begin{equation}
\label{chidot2} 
c_{e\!f\!f} \,\dot\chi = {1\over \tau_{pl}}\,e^{- e_Z/\chi}\,e^{e_Z/\chi_{\infty}}\, s_y\, [[{s_0\over s_y} - 1]] \Bigl(1 - {\chi\over \chi_{\infty}}\Bigr).
\end{equation}
Then, by equating coefficients of $\theta^2$ in Eq.(\ref{chidot}), we obtain an equation of motion for the new angular variable $\gamma(t)$:
\begin{eqnarray}
\label{gammadot}
\nonumber &&c_{e\!f\!f} \,\dot\gamma = {1\over \tau_{pl}}\, [[{s_0\over s_y} - 1]]\, e^{- e_Z/\chi}\,e^{e_Z/\chi_{\infty}}~~~~~~\cr\\&&\times\Biggl[4 \bigl(1- {\chi\over\chi_{\infty}}\bigr) + \Bigl(\bigl(1- {\chi\over\chi_{\infty}}\bigr){e_Z\over \chi^2} - {1\over \chi_{\infty}}\Bigr)\gamma\Biggr].~~~~
\end{eqnarray}

As in earlier papers, it is convenient to introduce the notation $\tilde\chi = \chi/e_Z$ and $\tilde \gamma = \gamma/e_Z$.  Making this substitution, and transforming to the dimensionless variables introduced in Sec.\ref{Bingham2}, I find for the tip displacement: 
\begin{equation}
\label{utip2}
{d \tilde u_{tip}\over d\psi} = {\bar\nu -1\over 2\,\kappa} \Bigl(c_0 {d \tilde s_0\over d\psi} + \xi \,\Gamma(\tilde \chi)[[\tilde s_0 -1]]\Bigr) + {c_0\over 2 \bar\nu^2 \sqrt{\kappa}}.
\end{equation}
where 
\begin{equation}
\label{Gammadef}
\Gamma(\tilde \chi) \equiv e^{- 1/\tilde\chi+1/\tilde\chi_{\infty}}.
\end{equation}
Equation (\ref{utip2}) is the same as Eq.(\ref{utip}) except for the factor $\Gamma(\tilde \chi)$ multiplying $\xi$.  Similarly the tip-stress equation (\ref{tipstress}) is
\begin{equation}
\label{tipstress2}
{d \tilde s_0\over d\psi} = - {\xi\over c_0}\Gamma(\tilde \chi) [[\tilde s_0 -1]] + \sqrt{\kappa}\, \Lambda(\bar\nu),
\end{equation}

\begin{figure}[h]
\begin{center}
\includegraphics[width=\linewidth] {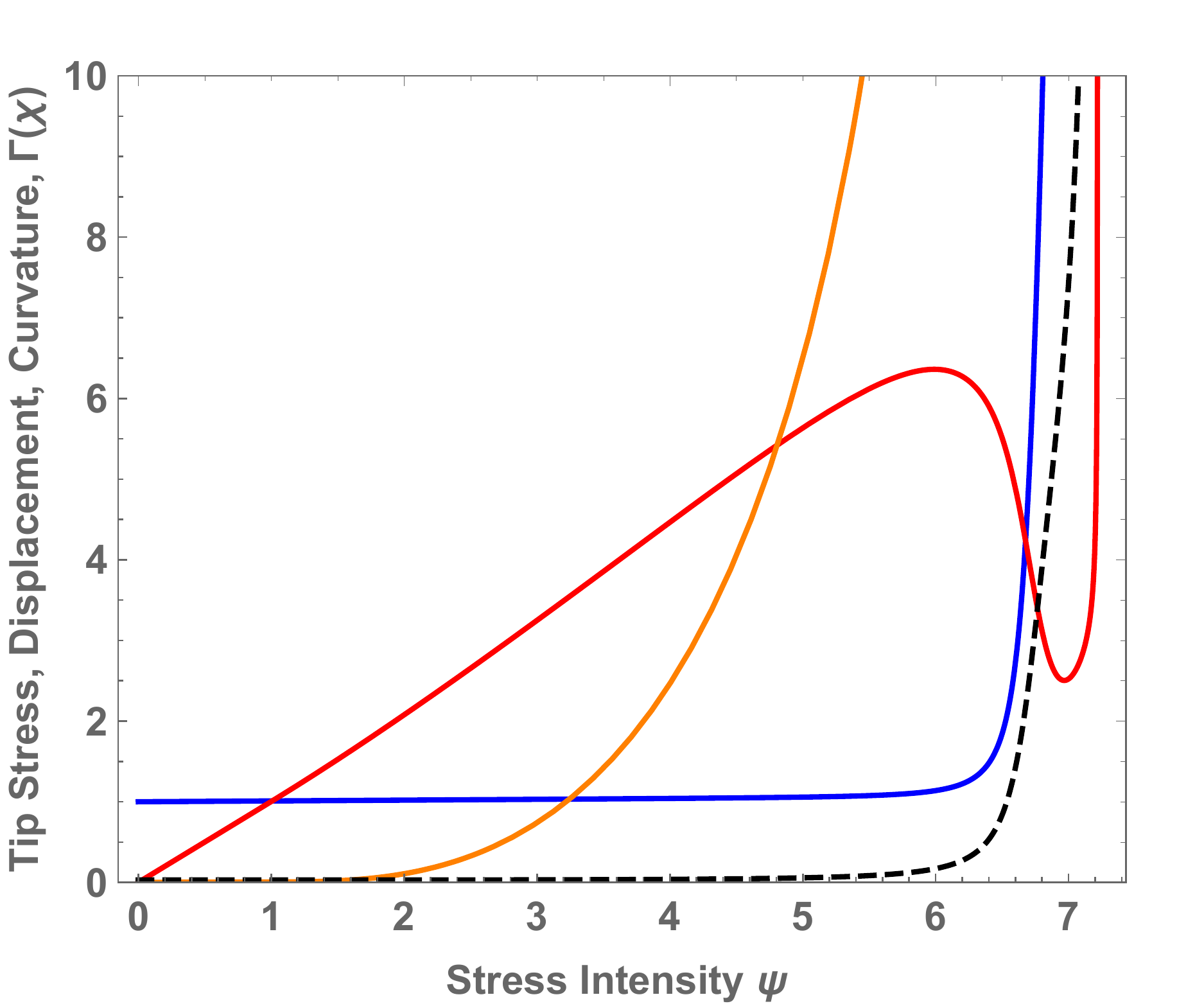}
\caption{Effective-temperature model with $\xi = 2$ and $T_0^{e\!f\!f} = 640\,K$.  The curves show the dimensionless tip stress $\tilde s_0$ (red), the curvature $\kappa$ (blue), and the displacement $\tilde u_{tip}$ (orange dashed) as functions of the stress intensity factor $\psi$. The black dashed curve is the STZ density factor $\Gamma(\chi)$ multiplied by a factor of $100$ for visibility. }   \label{BD3-Fig-5}
 \end{center}
\end{figure}

The curvature equation becomes
\begin{eqnarray}
\label{kappadot6}
\nonumber
&&{1\over \kappa}{d \kappa\over d\psi} = {(\bar\nu -1)^2\over 3} \Bigl(c_0 {d \tilde s_0\over d\psi} + \xi \,\Gamma(\tilde\chi)\,[[\tilde s_0 -1]]\Bigr)~~~~~\cr\\&&+ (\bar\nu -1)\,\xi \,\Gamma(\tilde\chi)\,[[\tilde s_0 -1]]\,{\tilde\gamma\over \tilde\chi^2} + 2 c_0 \sqrt{\kappa}\Bigl({2 \bar\nu -1\over  \bar\nu^2 }\Bigr)~~~~~~~
\end{eqnarray}.
The equation of motion for $\tilde\chi$, Eq.(\ref{chidot2}), becomes
\begin{equation}
\label{chidot3} 
{d\tilde\chi\over d\psi} =c_1\, \xi \,\Gamma(\tilde\chi) \,[[\tilde s_0 -1]]\,\Bigl(1 - {\tilde\chi\over \tilde\chi_{\infty}}\Bigr),
\end{equation}
where $c_1 = (s_y/ e_Z\,c_{e\!f\!f})$ is a dimensionless prefactor. Finally, the equation of motion for $\tilde\gamma$, Eq.(\ref{gammadot}), becomes
\begin{eqnarray}
\label{gammadot2}
\nonumber &&{d\tilde\gamma \over d\psi}= c_1\,\xi\,\Gamma(\tilde\chi)\, [[\tilde s_0 -1]] ~~~~~~\cr\\&&\times\Biggl[4 \bigl(1- {\tilde\chi\over\tilde\chi_{\infty}}\bigr) + \Bigl(\bigl(1- {\tilde\chi\over\tilde\chi_{\infty}}\bigr){1\over \tilde\chi^2} - {1\over \tilde\chi_{\infty}}\Bigr)\tilde\gamma\Biggr].~~~~
\end{eqnarray}
\begin{figure}[h]
\begin{center}
\includegraphics[width=\linewidth] {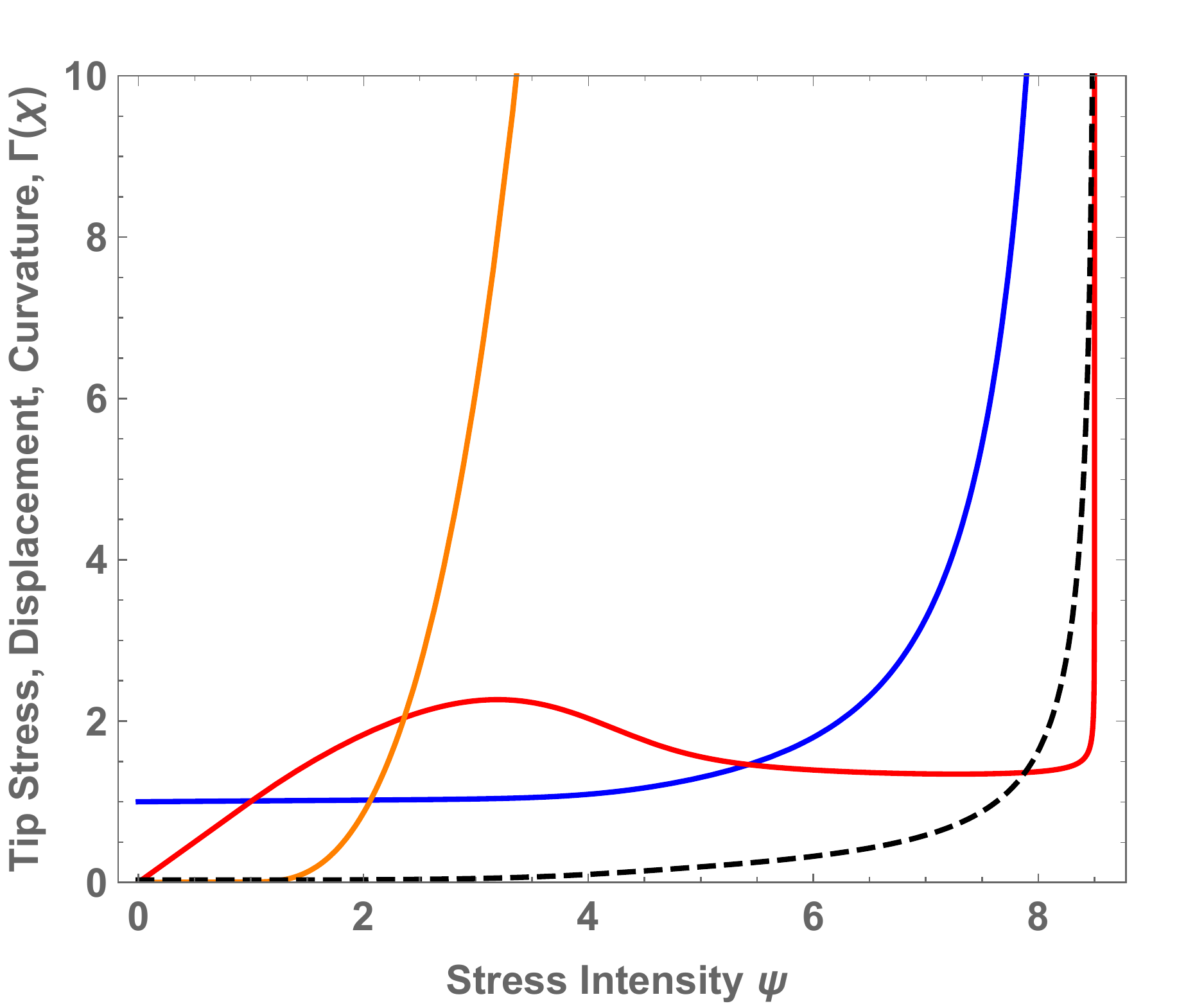}
\caption{Effective-temperature model with $\xi = 20$ and $T_0^{e\!f\!f} = 640\,K$.  The curves are the dimensionless tip stress $\tilde s_0$ (red), the curvature $\kappa$ (blue), and the displacement $\tilde u_{tip}$ (orange dashed) as functions of the  stress intensity factor $\psi$. The black dashed curve is the STZ density factor $\Gamma(\chi)$ multiplied by a factor of $100$ for visibility.}   \label{BD3-Fig-6}
 \end{center}
\end{figure}
\section{Numerical Results for the Effective-Temperature Model}
\label{Numerics}

To solve Eqs.(\ref{utip2}) - (\ref{gammadot2}) and compare the results with the numerical simulations of \cite{RB-12} and experimental data from \cite{SCHetal-18}, we need only a small number of system parameters specific to Vitreloy 1. I already have noted in Sec.\ref{Bingham2} that $c_0 = s_y/2\,\mu \cong 0.01$.  Here we also need $c_1 \cong 0.1$, a value that I deduce from \cite{VRB-16}.  As will be seen, this fairly small value of $c_1$ means that the sharp increase in the STZ density does not occur until the final plasticity-dominated phase of ductile fracture initiation. 

\begin{figure}[h]
\begin{center}
\includegraphics[width=\linewidth] {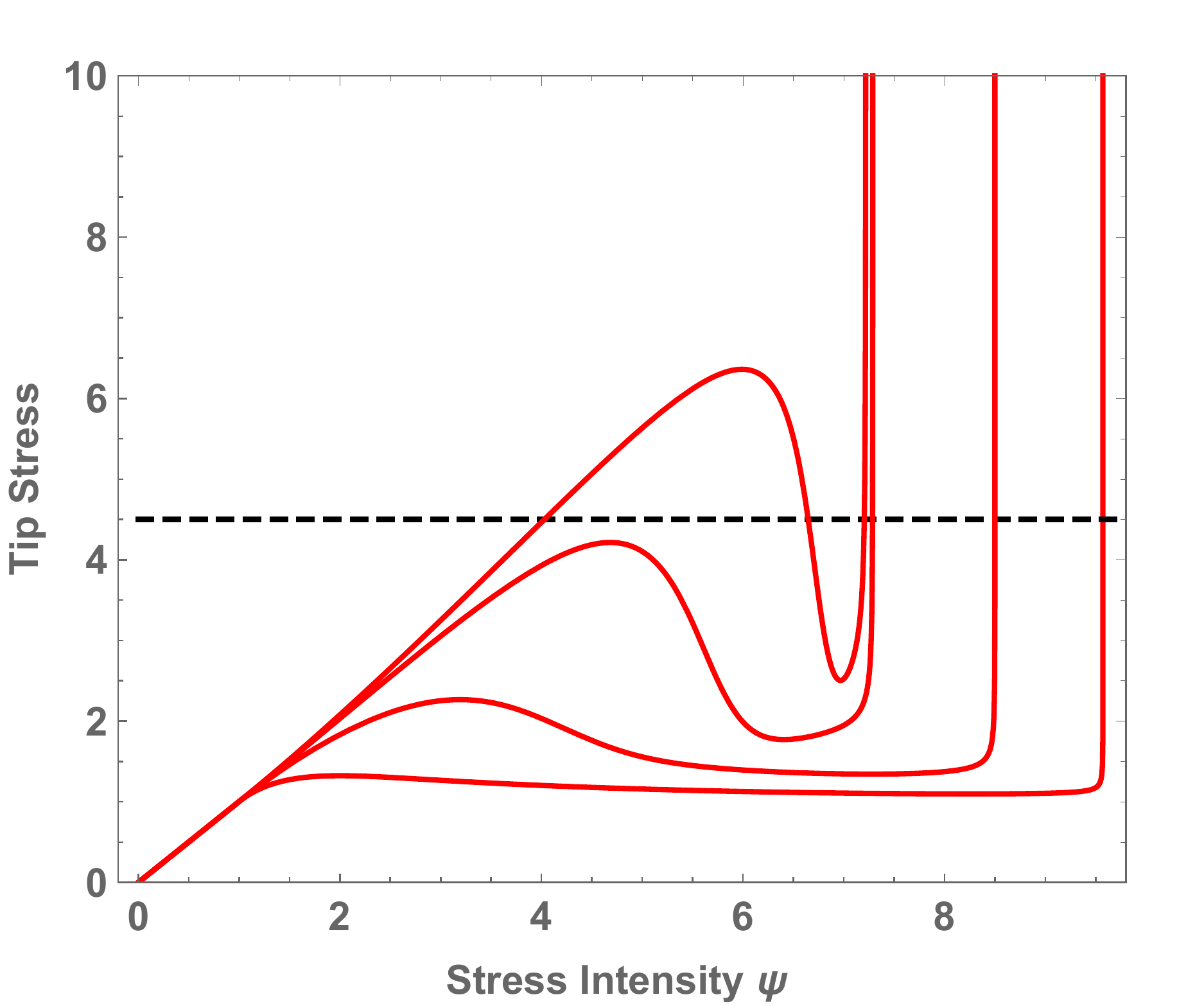}
\caption{Effective-temperature model.  The curves are the dimensionless tip stresses $\tilde s_0$ as functions of the stress intensity factor $\psi$ for $\xi = 2,\,5,\,20,\,{\rm and}\,\,100$ from top to bottom, for $T_0^{e\!f\!f} = 640\,K$. The horizontal line is at  $\tilde s_c = 4.5$.}   \label{BD3-Fig-7}
 \end{center}
\end{figure}

As in \cite{RB-12}, I denote values of $\tilde \chi$ by effective  temperatures.  Thus $\tilde\chi_{\infty} \equiv k_B T_{\infty}^{e\!f\!f}/ e_Z$ and, according to \cite{RB-12}, $T_{\infty}^{e\!f\!f} \cong 900\,K$.  Similarly, initial effective temperatures are denoted by $T_0^{e\!f\!f}$ with $\tilde\chi_0 =k_B T_0^{e\!f\!f}/ e_Z$.

The conversion from experimental units of fracture toughness to values of the dimensionless variable $\psi$ is easily accomplished just by fitting the single toughness value in the elastic limit at small $\xi$.  Thus, I find that my values of fracture toughness are approximately the reported values $K_Q$ in units of ${\rm MPa}\,\sqrt{m}$ multiplied by a factor $0.06$.  Similarly, the conversion from driving rate $\dot K_I$ to $\xi$ needs only a single fitting parameter, $\xi \cong 320/\dot K_I$.   

Figures \ref{BD3-Fig-5} and \ref{BD3-Fig-6} are analogous to Figs. \ref{BD3-Fig-1} and \ref{BD3-Fig-2} in that they show $\tilde s_0$, $\tilde u_{tip}$, and $\kappa$ as functions of $\psi$.  They also show graphs of the STZ density factor $\Gamma(\chi)$, defined in Eq.(\ref{Gammadef}), multiplied in the figure by a factor of $100$ for visibility. Both of these figures are computed with an initial effective temperature $T_0^{e\!f\!f} = 640\,K$. Figure \ref{BD3-Fig-5} is plotted for a relatively small rate ratio, $\xi =2$.  The peak in the tip stress looks much like the peak in Figure 2 of \cite{VRB-16} which was obtained via a circle approximation roughly similar to the one used here but without the boundary-layer dynamics or the relation to the tip parameters.  Here, the top of the peak has no special significance; a brittle crack would have been launched earlier when the stress crossed the critical value of $\tilde s_c = 4.5$. 

Figure \ref{BD3-Fig-6}, for $\xi = 20$, shows what happens on the ductile side of the transition.  The peak in $\tilde s_0(\psi)$ at $\psi \cong 5$ has dropped below $\tilde s_c$ because of plastic shielding of the notch tip and, thus, the notch continues to elongate and sharpen until it reaches the ductile failure limit at $\psi \cong 8.5$.  

Figure \ref{BD3-Fig-7} shows a set of $\tilde s_0(\psi)$ curves, analogous to those shown for the Bingham model in Fig. \ref{BD3-Fig-3}, here for $\xi = 2,\,5,\,20,\,{\rm and}\,\,100$ from left to right.  Apparently, an abrupt brittle to ductile transition occurs for $\xi$ just slightly smaller than $5$, where the $\tilde s_0(\psi)$ curve is tangent to the horizontal line at $\tilde s_c = 4.5$.  Above that value of $\xi$, ductile failure occurs when  $\tilde s_0(\psi) = \tilde s_c$ at larger values of $\psi$ where the plastic zone expands suddenly and the notch tip is no longer shielded. 

\begin{figure}[h]
\begin{center}
\includegraphics[width=\linewidth] {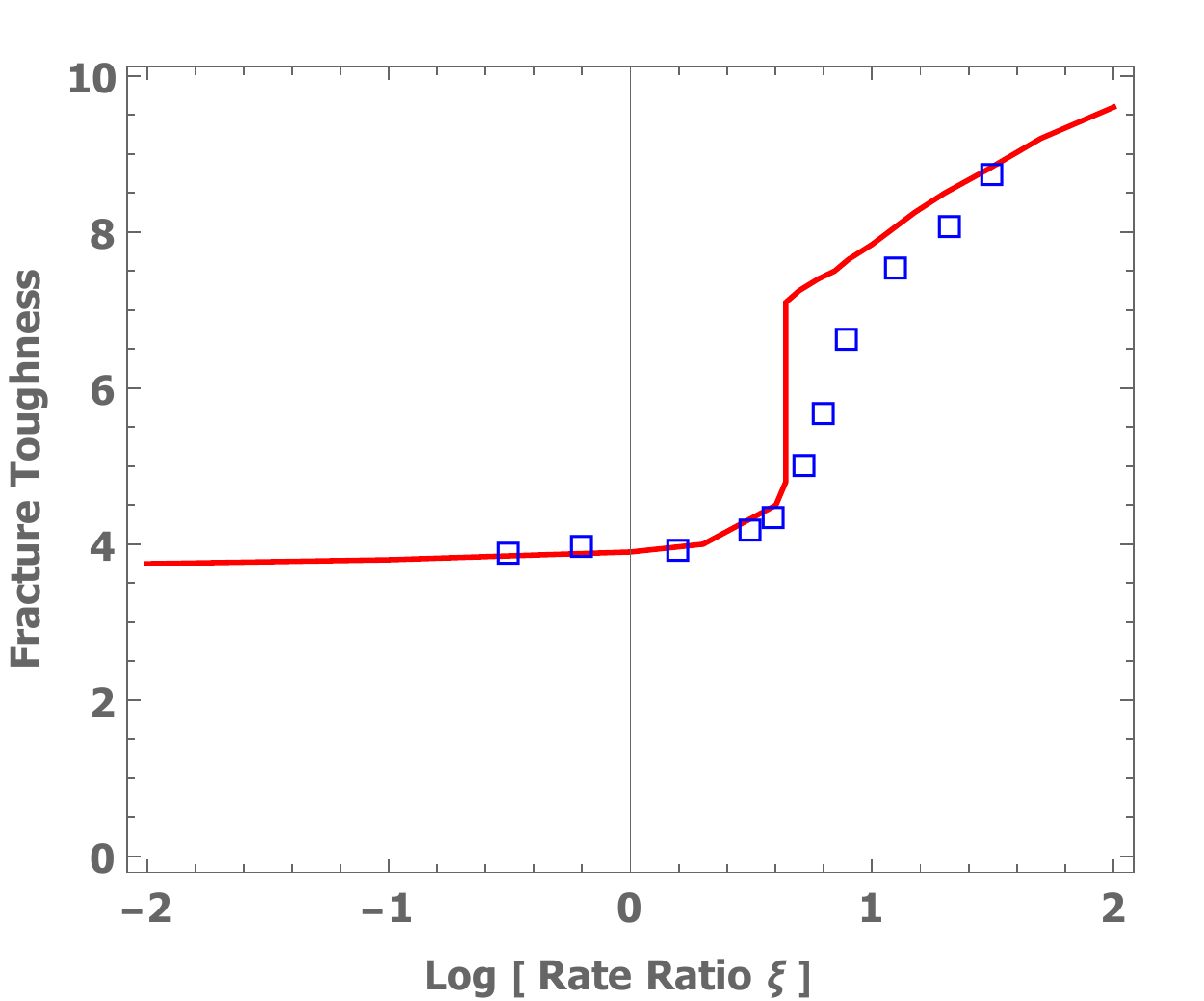} 
\caption{Comparison of the effective-temperature theory predictions of fracture toughness with the numerical simulation results in \cite{VRB-16} for $T_0^{e\!f\!f} = 640\,K$.  The effective-temperature results are shown by the solid red curve; the simulation results are the squares.}   \label{BD3-Fig-8}
 \end{center}
\end{figure}

\begin{figure}[h]
\begin{center}
\includegraphics[width=\linewidth] {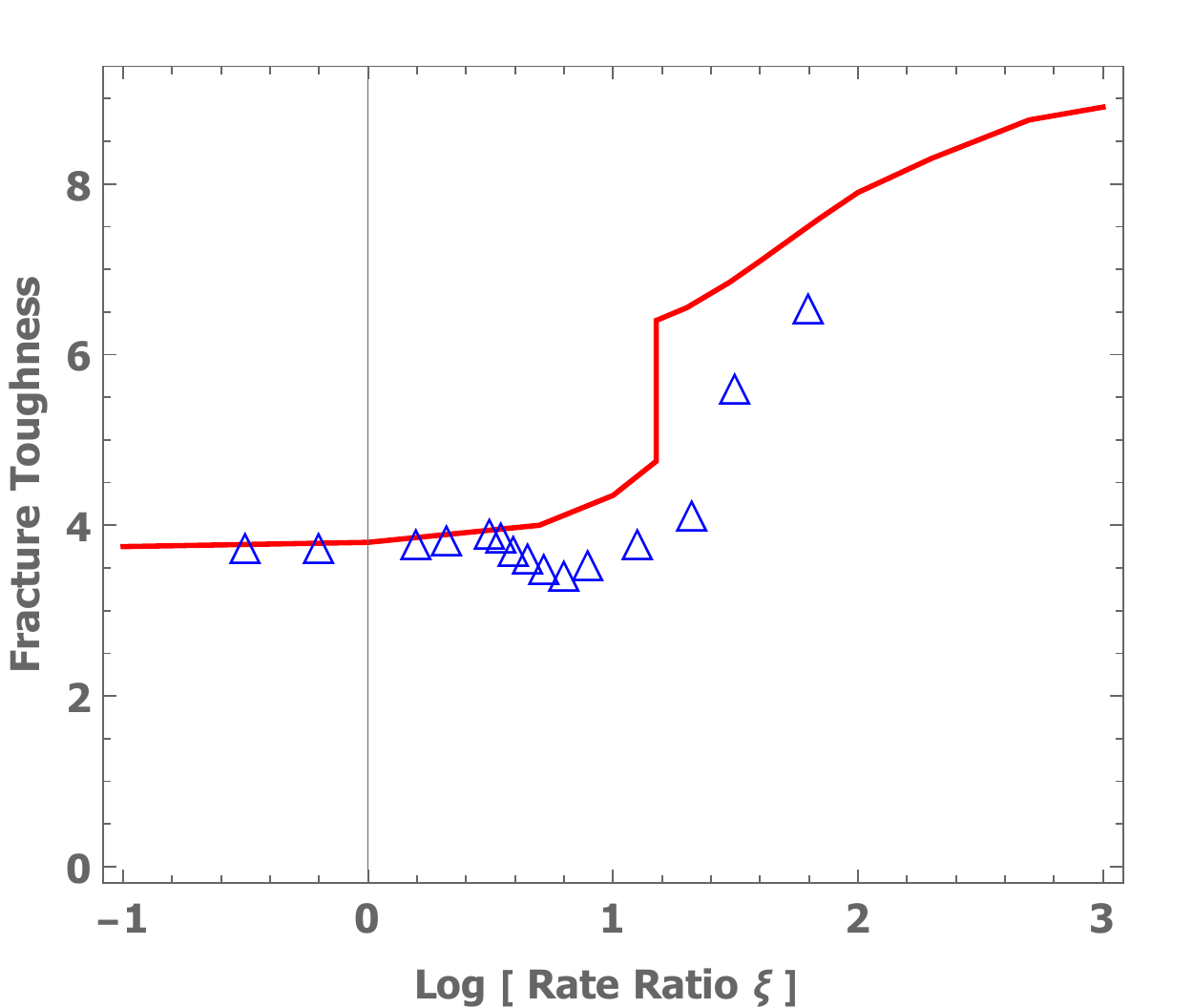}
\caption{Comparison of the effective-temperature theory prediction of fracture toughness as a function of the rate ratio $\xi$ (red curve) with the numerical simulation results taken from  \cite{VRB-16} (triangles).  All data are for $T_0^{e\!f\!f} = 610\,K$. }   \label{BD3-Fig-9}
 \end{center}
\end{figure}

Figures \ref{BD3-Fig-8} and \ref{BD3-Fig-9}  show comparisons between predictions of the effective-temperature theory and the numerical simulation data shown in Fig. 4 of \cite{VRB-16}.  These two figures are drawn from the data for fracture toughness as functions of driving rate $\dot K_I$ for $T_0^{e\!f\!f} = 640\,K\,\, {\rm and}\,\,610\,K$ respectively. They translate into toughness as functions of the rate ratio $\xi$.  

\begin{figure}[h]
\begin{center}
\includegraphics[width=\linewidth] {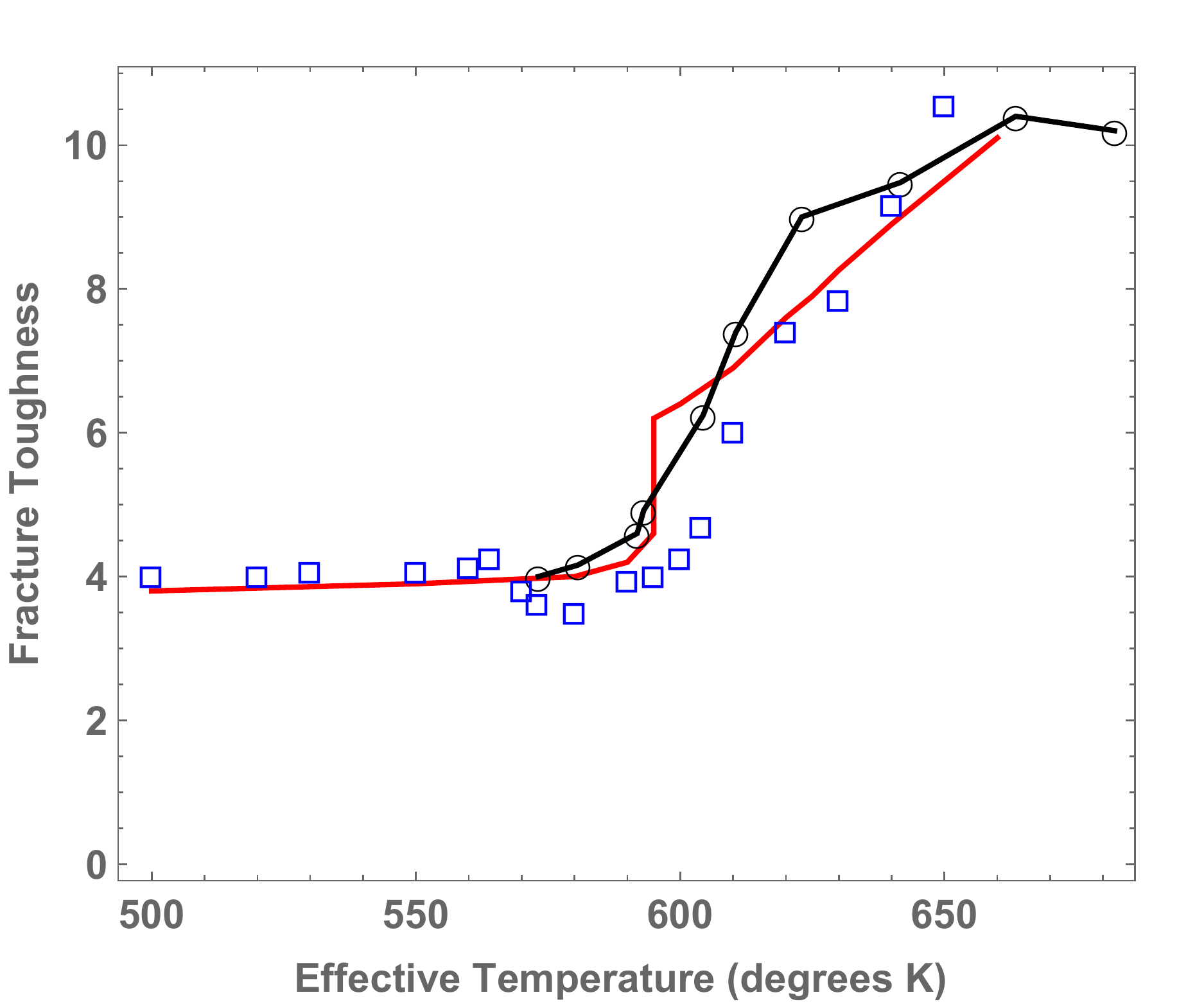}
\caption{Comparison of the effective-temperature theory prediction of fracture toughness as a function of $T_0^{e\!f\!f}$ (red curve) with the numerical simulation results taken from \cite{VRB-16} (blue squares) and experimental data taken from \cite{SCHetal-18} (joined black circles). }   \label{BD3-Fig-10}
 \end{center}
\end{figure}

Finally, Fig. \ref{BD3-Fig-10} shows toughness as a function of the initial effective temperature $T_0^{e\!f\!f}$ for a fixed driving rate, that is, for $\dot K_I =  10\,{\rm MPa}\,\sqrt{m}/s$ in Fig. 4 of \cite{VRB-16} (square data points), and theoretically for $\xi = 32$ (red curve). This is the one place where I can make a direct comparison with experimental data.  The joined circles in Fig. \ref{BD3-Fig-10} are taken from Fig. 1a in \cite{SCHetal-18}.  They  should be directly comparable with the other two data sets shown in  this figure.  

I find these admittedly rough comparisons to be both encouraging and thought provoking.  The agreement between theory, simulation, and experiment is good in the sense that the magnitudes and positions of the brittle-ductile transitions are well predicted without the use of arbitrary fitting parameters.  But there is obviously a substantial amount of uncertainty about all three data sets in Fig. \ref{BD3-Fig-10}; and whether the data rules out -- for example -- the discontinuity in the theory seems to me to be an open question. 

In my opinion, the fact that the predicted transitions are mathematically discontinuous means that the theory probably is unrealistic in some way.  Also, the clear non-monotonicity of the simulation data in Figs. \ref{BD3-Fig-9} and \ref{BD3-Fig-10} seems  interesting, as has been pointed out by the authors of \cite{VRB-16}.  But there is no hint of that non-monotonicity in the experiments.  

I suspect that the main missing ingredient in this theory is a sufficiently detailed description of the changing shape of the notch tip.  For example, Figure 2b in \cite{RB-12} shows a bulge with a radius of curvature roughly half that of the tip emerging from the front of the notch and substantially raising the effective temperature in its neighborhood.  Rycroft has shown unpublished movies of simulated later stages of ductile yielding in which a bulge of that kind moves forward for a considerable distance before launching a fast crack. There are only hints of such behavior in the present theory.   Note that the theoretical notch tip described by the graphs in Fig. \ref{BD3-Fig-6} does sharpen before reaching its failure limit.

\section{Remarks}
\label{Questions}

I have long been skeptical about various aspects of conventional materials science, especially dislocation theory (e.g. see \cite{JSL-19}), because results often are based on non-predictive phenomenology.  The results presented here make me more optimistic about opportunities for improving the situation.  There are many open issues.

{\it Yielding Transitions.} A key assumption throughout this analysis is that the plastic yielding transition is sharp and nonsingular, {\it i.e.} that it is Bingham-like near threshold.  This is not true in many rheological models, for example, in Herschel-Bulkley models where the flow rate is proportional to the square root of the incremental stress above the yield stress $s_y$.  There is also the possibility that yielding transitions may be critical phenomena accompanied by diverging fluctuations.  That is what happens in athermal quasistatic models which ignore the fact that internal relaxation rates necessarily become faster than external driving rates when the latter vanish at a yield point. 

My papers \cite{JSL-08} and \cite{JSL-15a} were written in large part to explore the nature of yielding transitions in STZ theories of amorphous materials, especially metallic glasses.   In \cite{JSL-15a}, I argued from first principles that realistic yielding transitions of this kind are non-critical.  I also showed in \cite{JSL-15a} that the Bingham model can be derived as a limit of STZ theory.

{\it Dislocations.} One of my original motives for starting this project was the idea that the new thermodynamic dislocation theory \cite{LBL-10,JSL-17a,JSL-19} must be relevant to fracture toughness.  The problem of understanding brittleness and ductility in metals and alloys and other crystalline materials is far more complex than it is for amorphous materials.  Just the existence of multiple slip systems and grain boundaries and the like makes this topic seem formidable.  Nevertheless, important progress has been made in the last decade simply by realizing that dislocations in driven systems must obey the second law of thermodynamics and thus must be amenable to an effective-temperature analysis.  That realization has led to successful first-principles theories of strain hardening and sharp yielding transitions, both of which are relevant to fracture.  

The picture developed here of brittle fracture being initiated in a metallic glass at a low fictive temperature looks almost identical to the picture of a notch in an unhardened crystalline material with a low initial density of dislocations.  The external stress generates dislocations at an effectively hot spot at the tip of the notch.  It should be possible to use the new dislocation theory to predict how rapidly that happens and what happens next.  There are many such opportunities for progress along these lines.

{\it Fracture Dynamics.} This theory of the onset of brittle or ductile fracture occupies just a tiny corner of the large field of fracture dynamics.  It is not obvious how to bridge the gap between this corner and the rest of the field.  

Note that my equations of motion in Secs.\,\ref{Bingham} and \ref{Teff} do not look like those that appear in the conventional literature on fracture dynamics.  In the conventional picture (e.g. see Freund \cite{FREUND}), we visualize a Griffiths-like crack moving on a well defined plane, driven by remote loading that causes elastic energy to flow to the crack tip where that energy is somehow dissipated.  For many years, the most promising  descriptions of the tip behavior seemed to be the cohesive-zone models of Dugdale and Barenblatt.\cite{DUGDALE,BARENBLATT}  Sometimes those cohesive zones were called ``plastic'' zones; but the models never included realistic equations of motion for the plastic flow fields $v_{\rho}^{pl}$ and $v_{\theta}^{pl}$ that appear here.  Moreover, it has been known for twenty years  that most cohesive-zone models are intrinsically ill-posed; they produce strongly unstable cracks if they describe cracks at all.\cite{CLN,monster}

In my opinion, some of the most interesting recent developments in fracture dynamics are those described by Bouchbinder and colleagues in Refs.\cite{BFM-10, BK-17, BK-18}. These authors develop nonlinear field theories to describe the dynamics of fast cracks, and show that their theories predict high-speed behaviors, including instabilities and sidebranching, in agreement with experimental observations.  Those theories are not -- and cannot be -- simple extensions of the quasistatic onset behavior studied here.  Both of these related but qualitatively different classes of behavior  -- the onset behavior that determines brittleness and ductility, the late-stage behavior that determines large-scale failure, and the range of phenomena that lies between them -- continue to be highly promising areas for research.

\appendix
\section{Elliptical Formulas}

For completeness, I list in this Appendix the formulas on which I have based my analyses.  The elliptical coordinates are defined in Eq.(\ref{rhoW}).  

First, there are expressions for the rate-of-deformation tensor   ${\cal D}$ in terms of the elliptical material velocity components $v_{\rho}$ and $v_{\theta}$ as derived from more general formulas in Malvern \cite{MALVERN}.
\begin{equation}
\label{Drhorho}
{\cal D}_{\rho\rho} = {1\over WN}\,\left[{\partial v_{\rho}\over \partial\rho}+{v_{\theta}\over\rho}\,{1\over N}\, {\partial N\over\partial\theta}\right];
\end{equation}
\begin{equation}
\label{Dthetatheta}
{\cal D}_{\theta\theta}={1\over WN\rho}\,\left[{\partial v_{\theta}\over \partial\theta}+{v_{\rho}\over N}\, {\partial \over\partial\rho}(\rho N)\right];
\end{equation}
where the metric function is
\begin{equation}
\label{Nmetric}
N^2(\rho,\theta)= 1+{m^2\over\rho^4}-{2m\over\rho^2}\,\cos 2\theta.
\end{equation} 

Then there are the formulas for incompressible, two-dimensional elasticity that I have derived from Mushkelishvili \cite{MUSK-63}.  The following formulas assume vanishing normal stress on the surface of the elliptical hole, that is, at $\rho = 1$.  The stress tensor $\sigma$ is given by
\begin{equation}
\label{sigmaeqn1}
\sigma_{\rho\rho}+\sigma_{\theta\theta}=\sigma_{\infty}\, Re\left[1+ {2(1+m)\,e^{-2i\theta}\over \rho^2-m\,e^{-2i\theta}}\right];
\end{equation}
and
\begin{eqnarray}
\label{sigmaeqn2}
\nonumber
&&{\cal S}(\rho,\theta)\equiv \sigma_{\theta\theta}-\sigma_{\rho\rho}+2i\sigma_{\rho\theta}~~~~ \cr\\&&=\nonumber{\sigma_{\infty}\rho^2 e^{2i\theta}\over \left(\rho^2-m\,e^{2i\theta}\right)}\cr\\&& \times\left[1-{e^{-2i\theta}\over m\rho^2}+{(1+m)\,e^{-2i\theta}\over \left(\rho^2-m\,e^{-2i\theta}\right)^2} \,M(\rho,\theta)\right]
\end{eqnarray}
where
\begin{eqnarray}
\label{Mtheta}
\nonumber
M(\rho,\theta)&&={\rho^2\over m}\left(1-2\,m\,e^{-2i\theta}+m^2\right)\cr\\&& + e^{i\theta}\left(1-2\,m\,e^{2i\theta}+m^2\right).
\end{eqnarray}
According to (\ref{sigmaeqn2}) the deviatoric stress has components
\begin{equation}
\label{devstrss}
s_{\theta\theta} =-s_{\rho\rho}={1\over 2}\, Re\,{\cal S}(\rho,\theta);~~~~s_{\rho\theta}={1\over 2}\, Im\,{\cal S}(\rho,\theta).
\end{equation}

\begin{acknowledgments}

JSL was supported in part by the U.S. Department of Energy, Office of Basic Energy Sciences, Materials Science and Engineering Division, DE-AC05-00OR-22725, through a subcontract from Oak Ridge National Laboratory.  

\end{acknowledgments}

\end{document}